\documentclass{aa}  

\usepackage{graphicx}
\usepackage{xcolor}
\usepackage{adjustbox}
\usepackage[load-configurations = abbreviations]{siunitx}

\usepackage{txfonts}
\begin{document} 

   \title{Flattened loose particles from numerical simulations 
compared to Rosetta collected particles}


   \author{J. Lasue\inst{1}
          \and
          I. Maroger\inst{1}
          \and
          R. Botet\inst{2}
	\and
          Ph. Garnier\inst{1}
	\and
          S. Merouane\inst{3}
	\and
          Th. Mannel\inst{4,5}
	\and
          A.C. Levasseur-Regourd\inst{6}
	\and
          M.S. Bentley\inst{7}
          }

   \institute{IRAP, Université de Toulouse, CNRS, CNES, UPS, Toulouse, France\\
             \email{jlasue@irap.omp.eu}
         \and
            Université Paris-Saclay/Université Paris-Sud/CNRS, UMR 8502, LPS, Orsay, France
         \and
            Max Planck Institute for Solar System Research, Göttingen, Germany
         \and
            Space Research Institute of the Austrian Academy of Sciences, Graz, Austria
         \and
            University of Graz, Graz, Austria
         \and
            Sorbonne Université, CNRS, LATMOS, Paris, France
         \and
            European Space Astronomy Centre, Madrid, Spain
             }

   \date{Received September 15, 1996; accepted March 16, 1997}

 
  \abstract
   {Cometary dust particles are remnants of the primordial accretion of refractory material 
   that occurred during the initial stages of the Solar System formation.
   Understanding their physical structure can help constrain their 
   accretion process.}
   {The {\it in situ} study of dust particles collected at slow speeds 
   by instruments on-board the Rosetta space mission, including 
   GIADA, MIDAS and COSIMA, can be used to infer 
   the physical properties, size distribution, and typologies of the dust.}
   {We have developed a simple numerical simulation of aggregate impact flattening 
   to interpret the properties of particles collected by COSIMA. 
   The aspect ratios of flattened particles from both simulations and observations 
   are compared to differentiate between initial families of aggregates 
   characterized by different fractal dimensions $D_f$. 
   This dimension can differentiate between certain growth modes, namely 
   the Diffusion Limited Cluster-cluster Aggregates (DLCA, $D_f \approx 1.8$), 
   Diffusion Limited Particle-cluster Aggregates (DLPA, $D_f \approx 2.5$), 
   Reaction Limited Cluster-cluster Aggregates (RLCA, $D_f \approx 2.1$), and 
   Reaction Limited Particle-cluster Aggregates (RLPA, $D_f \approx 3.0$).}
   {The diversity of aspect ratios measured by COSIMA is consistent with either
   two families of aggregates with different initial $D_f$ 
   (a family of compact aggregates with fractal dimensions close to 2.5-3 
  and some fluffier aggregates with fractal dimensions around 2). 
  Alternatively, the distribution of morphologies seen by COSIMA could originate from 
  a single type of aggregation process, such as DLPA, 
  but to explain the range of aspect ratios observed by COSIMA 
  a large range of dust particle cohesive strength is necessary.
  Furthermore, variations in cohesive strength and velocity may play a role in 
  the higher aspect ratio range detected (>0.3).
  }
   { Our work allows us to explain the particle morphologies observed by 
  COSIMA and those generated by laboratory experiments 
  in a consistent framework. 
  Taking into account all observations from the three dust 
  instruments on-board Rosetta, we favor an interpretation of our 
  simulations based on two different families  of dust 
  particles with significantly distinct fractal dimensions ejected from 
  the cometary nucleus.}


\titlerunning{Flattened loose particles compared to Rosetta}
\maketitle
%

\section{Introduction}

\subsection{Cometary dust particles}

  Comets are believed to preserve pristine dust 
  grains and to provide information about their aggregation 
  processes in the early Solar System 
  \citep[e.g.][]{weidenschilling_origin_1997, blum_laboratory_2000}.
  Analyses of data from the Giotto mission to comet 
  1P/Halley and of foil impacts and aerogel tracks  retrieved 
  by the Stardust mission in the coma of comet 81P/Wild 2 
  have indeed given clues to the presence of low density dust 
  particles built up of agglomerates, possibly with different tensile 
  strengths and porosities \citep[e.g.][]{fulle_situ_2000, horz_impact_2006, 
  burchell_characteristics_2008}.
  The interpretation of remote polarimetric observations of bright 
  comets, such as 1P/Halley and C/1995 O1 Hale-Bopp, has 
  lead to similar conclusions 
  \citep{levasseur-regourd_dust_2008, lasue_cometary_2009}.
  Aggregation of solid particles in the early Solar System 
  may therefore form a diversity of porosities 
  represented by their fractal dimension, $D_f$ 
  \citep{dominik_physics_1997, kempf_n-particle-simulations_1999, 
  bertini_influence_2009}. 
  Understanding the structure of cometary 
  dust particles can give clues to these early Solar 
  System processes \citep{blum_growth_2008, fulle_fractal_2017}. 


\subsection{The Rosetta mission}

  During its 26 month long rendezvous with comet 
  67P/Churyumov-Gerasimenko (hereafter 67P) in its 2015 apparition, 
  the Rosetta spacecraft monitored the properties of cometary dust particles 
  released by the nucleus in the pre- and post- perihelion 
  phases, as well as during some outburst events. Three 
  instruments were specifically devoted to the study of dust 
  particles: i) COSIMA (the COmetary Secondary Ion Mass Analyzer, 
  \citet{kissel_cosimahigh_2007}) collected dust 
  particles of 10 to \SI{100}{\micro\meter} size 
  on \SI{1}{\centi\meter\squared} targets, imaged them with a microscope 
  operating under grazing incidence illumination with a resolution 
  of about \SI{14}{\micro\meter}, and then analyzed them 
  through a mass spectrometer after indium ion beam ablation,
  ii) MIDAS (the Micro-Imaging 
  Dust Analysis System, \citet{riedler_midasmicro-imaging_2007})
   collected micron-sized dust particles on targets of about 
  \SI{3.5}{\milli\meter\squared}, 
  in order to obtain 3D images of their surfaces down to 
  tens of nanometers pixel resolution using atomic
  force microscopy, and 
  iii) GIADA (the Grain Impact Analyzer and Dust Accumulator, 
  \citet{colangeli_grain_2007}) measured 
  the optical cross-section, speed, momentum and cumulative 
  flux of hundreds of sub-millimeter sized dust particles. 

  The COSIMA and MIDAS instruments collected dust particles 
  at velocities in the 1 to \SI{15}{\meter\per\second} range \citep{fulle_density_2015}, 
  that is to say at relative velocities much lower than the \SI{6.1}{\kilo\meter\per\second} 
  reached during the collection of 81P/Wild 2 samples. Their chemical 
  properties were thus mostly preserved, as well as part of their physical 
  structure. Some small particles, which could be fragments of fragile 
  individual particles, were nevertheless noticed 
  \citep[e.g.][]{bentley_aggregate_2016, merouane_dust_2016}.
  Interestingly enough, some particles appeared to be flattened, 
  most likely as a result of impact alteration 
  \citep[e.g.][]{langevin_typology_2016, mannel_fractal_2016}.

  The Rosetta dust experiments provide complementary insights into 
  the properties of dust particles thanks to their different approaches
  \citep[see, for a review,][]{levasseur-regourd_cometary_2018}.
  As far as images are concerned, the total number of dust particles
  detected is above 30,000 for COSIMA and above 1,000 for MIDAS
  \citep{levasseur-regourd_cometary_2018, Guttler_dust_2019}.

  More specifically, all images of dust particles indicate that they 
  consist of more or less porous agglomerates of smaller grains 
  (following the classification introduced in \citep{Guttler_dust_2019}).
  Their overall sizes, identified by well-defined boundaries, range 
  from about 1 micrometer to tens of micrometers for 
  MIDAS, and from tens of micrometers to several hundreds of micrometers 
  for COSIMA. The presence of aggregated structures at distinct 
  scales suggests a hierarchical aggregation \citep{bentley_aggregate_2016}. 
  Indeed, the fractal dimension of a very porous agglomerate detected 
  by MIDAS was determined via a density-correlation function 
  \citep{mannel_fractal_2016}, to be equal to 1.7$\pm$0.1. 
  Dust showers observed by GIADA were also explained by 
  the presence of fragile agglomerates with a fractal dimension below 2, possibly disrupted
  through electrostatic fragmentation induced by the spacecraft  
  \citep{fulle_density_2015, fulle_evolution_2016}.
  Considering fractal aggregation processes, the porosity of dust 
  particles in 67P/Churyumov-Gerasimenko can thus be estimated 
  to be at least equal to 90\% for very porous ones, and about 75\% for 
  more compact ones \citep[e.g.][]{blum_growth_2008, bertini_influence_2009}.
  The porosity of 67P/Churyumov-Gerasimenko's dust particles 
  has been estimated to be around 60\% based on the density 
  of the nucleus and the composition measured by COSIMA 
  \citep{fulle_dust--ices_2017}.
  Analysis of the reflectance of porous dust particles collected by COSIMA 
  indicate that a high porosity (>50\%) is necessary to explain that 
  the mean free path of photons in the particle correspond to 
  a significant fraction of the particle size \citep{langevin_optical_2017}.
  
  Finally, it may be added that the properties of cometary dust particles, 
  as revealed by the Rosetta mission, are, as previously suspected, 
  remarkably comparable to CP-IDPs, i.e. Chondritic Porous 
  Interplanetary Dust Particles collected in the Earth's stratosphere, 
  and UCAMMs, i.e. UltraCarbonaceous Antarctica MicroMeteorites 
  collected in the snows of central regions of Antarctica 
  \citep[e.g.][]{levasseur-regourd_cometary_2018}.
  
  The morphology, the structure and the composition of such dust particles 
  strongly suggest that, as well as cometary nuclei themselves, they formed in 
  the solar nebula and the primordial disk 
  \citep[e.g.][]{davidsson_primordial_2016, blumEvidenceFormationComet2017},
  and were never processed within large objects.

\begin{figure}
	\center
	\includegraphics[width = 0.45\columnwidth]{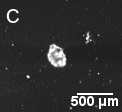}
	\includegraphics[width = 0.49\columnwidth]{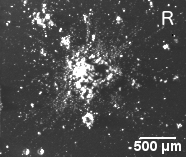}
	\includegraphics[width = 0.45\columnwidth]{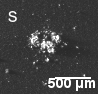}
	\includegraphics[width = 0.49\columnwidth]{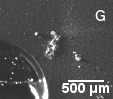}
	\caption{Diversity of crushed particle types detected by COSIMA 
	(Nick: compact particle, C; Alexandros: rubble pile, R; Estelle: shattered cluster, S; 
	and Johannes: glued cluster, G) (adapted from \citet{langevin_typology_2016}).}
	\label{fig:cosima}
\end{figure}

\subsection{Specificity of COSIMA results}

  COSIMA collected and analyzed cometary particles ejected by 
  67P/Churyumov-Gerasimenko on gold black covered targets 
  \citep{kissel_cosimahigh_2007}. The dust particles ejected 
  by the comet impacted COSIMA targets at a speed $<$\SI{10}{\meter\per\second} 
  according to GIADA measurements \citep{rotundi_dust_2015} 
  with a deceleration $<$\SI{1e6}{\meter\per\second\squared} 
  according to \citet{hornung_first_2016}. 
  These values are enough to damage 
  the initial structure of the dust particles during the collision, as 
  visually assessed from the images acquired by COSISCOPE after collection. 
  With a resolution of 14~microns, the microscope enabled studies of 
  particle typology and flux \citep{langevin_typology_2016}. 
  The images show particles ranging from a few tens to several 
  hundreds of microns, the majority of which appears to be built  
  of micron-sized sub-components, as confirmed by MIDAS 
  \citep{bentley_aggregate_2016}. Analysis of the particle morphologies 
  identified four families of particles \citep{langevin_typology_2016}
  which fall into two major classes, compact and clustered. These families are~:
\begin{enumerate}   
  \item{Compact (type C) particles present well-defined boundaries without smaller 
  satellite particles and with an apparent height above the collecting plane of the same order
  of magnitude as their horizontal ($x$ and $y$) dimensions.}
  \item{Shattered cluster (type S) particles are defined by clusters of fragments for 
  which no individual fragment makes up a major fraction of the initial particle. 
  These particles are interpreted as rearrangement of fragments within the 
  impacting particle without associated disruption.}
  \item{Glued cluster (type G) particles have a well-defined shape and a complex
  structure where sub-components appear to be linked by a fine-grained matrix
  with a smooth texture.}
  \item{Rubble piles (type R) particles comprise components much smaller 
  than their apparent size. Upon collision with the plate, the sub-components 
  rearranged themselves in a flattened conical pile with many satellite components 
  indicating poor cohesion.}  
\end{enumerate}  
  The different types of particles collected by COSIMA 
  are illustrated in Fig.~\ref{fig:cosima}.

\begin{figure}
	\center
	\includegraphics[width = \columnwidth]{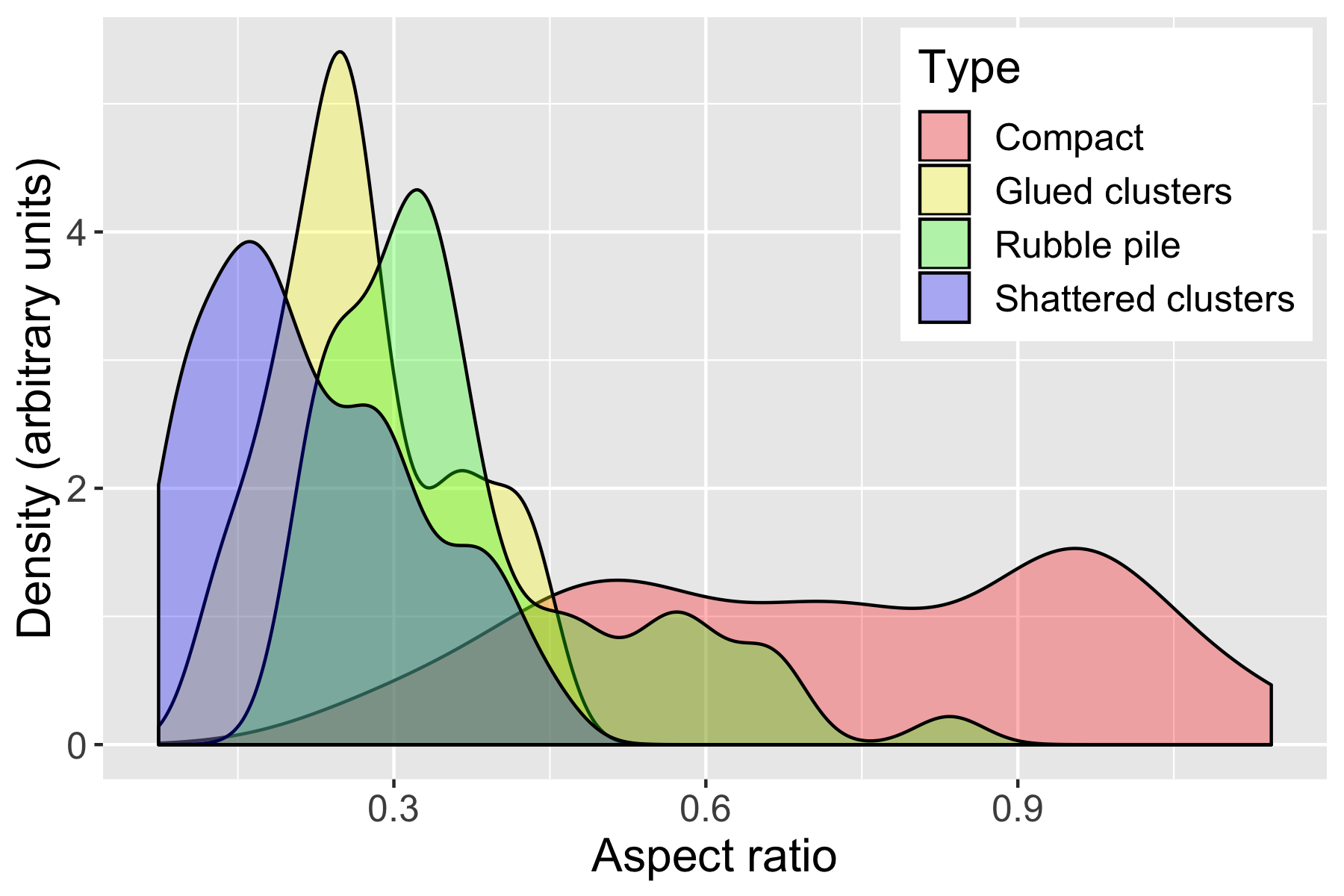}
	\caption{Probability density of aspect ratio for each type of particles detected by COSIMA 
	(adapted from \citet{langevin_typology_2016}).}
	\label{fig:langevin}
\end{figure}

  The grazing incidence illumination provided by COSISCOPE allows both 
  the surface area of the collected particles and their height (based 
  on their projected shadow) to be determined (see Fig.~\ref{fig:cosima}). 
  The area is determined from the ratio of bright pixels before and after 
  exposure to the dust flux from the comet. 
  An aspect ratio of the compacted particles can be obtained 
  from $\frac {height} {\sqrt {area}}$. The aspect ratio density distribution 
  for each detected particle type is shown in Fig.~\ref{fig:langevin}.  
  The compact particles, C, appear unbroken and present the largest 
  aspect ratios, with a first peak around 0.5 and another
  close to 1. The other particles present typical aspect ratios of around 0.3 with
  the shattered clusters being the flattest type of agglomerates.
  To understand the physical structure of cometary nuclei, it is important 
  to infer, as far as possible, properties of dust particles prior to their collection.  
  COSIMA analyses have shown a correlation between the flux of dust particles 
  at various distances from the comet nucleus and their morphology 
  \citep{merouane_dust_2016}. The fragmenting particles appear to have a 
  mechanical strength of a few \SI{1000}{\pascal} \citep{hornung_first_2016} 
  and their morphological diversity could result from different collection speeds
  in the range from \SIrange{1}{6}{\meter\per\second} 
  as investigated by laboratory simulations 
  \citep{ellerbroek_footprint_2017}. 

 In this work, we investigate if different dust particle structures 
 prior to their collection can also lead to the different morphologies 
 found by the Rosetta dust instruments.
 We present a set of numerical simulations of fractal aggregates
 flattening on impact with a plane surface, before presenting its  
 results and discussing their implications for the interpretation of 
 the Rosetta measurements. 

\section{Method}

\begin{figure}
	\center
	\includegraphics[width = \columnwidth]{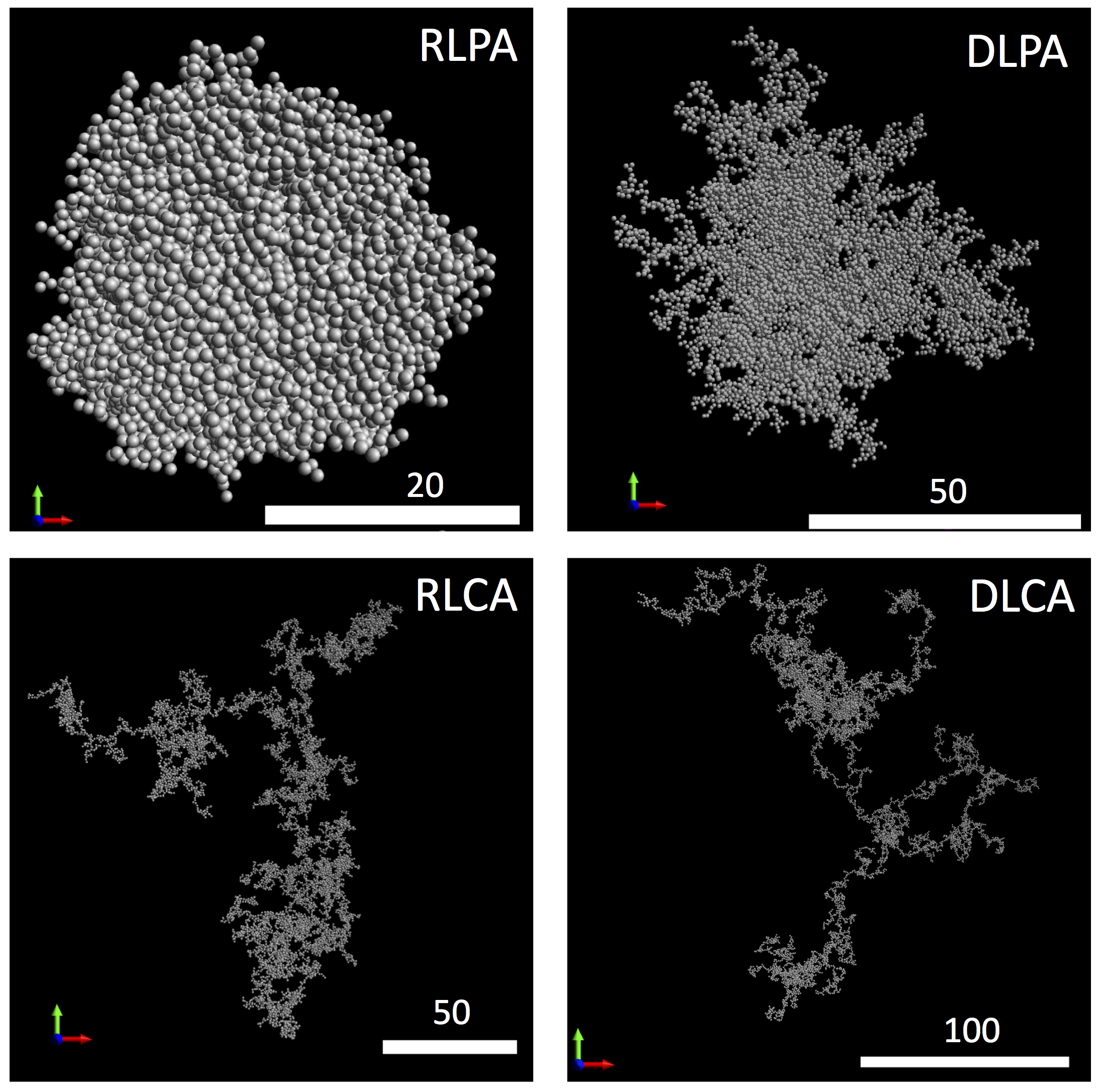}
	\caption{3D representation of four aggregates representing the four different aggregation processes
	considered in this work.
	DLCA ($D_f \approx 1.8$), RLCA ($D_f \approx 2.1$), 
         DLPA ($D_f \approx 2.5$) and RLPA ($D_f \approx 3$).
         A scale is given in number of monomers, and a referential frame 
         is indicated by colored arrows (x=blue, y=red, z=green).}
	\label{fig:aggregates}
\end{figure}

\subsection{Fractal aggregates models}

  We expect the dust particles aggregating in the solar nebula
  to present fractal structures. 
  Fractal aggregates in the early Solar System form a diversity of porosities 
  that can be represented by their fractal dimension, $D_f$, based on their 
  aggregation processes \citep{wurm_experiments_1998}. 
  Aggregation simulations consider the collisions of spherical monomers
  which represent individual grains aggregating to form dust particles \citep{Guttler_dust_2019}.
  Four main aggregation processes, leading to significantly different fractal 
  dimensions, are considered: DLCA ($D_f \approx 1.8$), 
  RLCA ($D_f \approx 2.1$), DLPA ($D_f \approx 2.5$) and RLPA ($D_f \approx 3$). 
  The DL models are Diffusion Limited models, in which when one monomer meets 
  another one it sticks directly to it. The RL models are Reaction Limited models, in which 
  molecular reactions occur when two monomers encounter each other 
  and result in them sliding with respect to one another in order to maximize the number of bonds,
  resulting in a more compact aggregate.
  CA stands for Cluster-cluster Aggregation and PA for Particle-cluster Aggregation :  
  in PA particles, monomers are added to the same main cluster which 
  accretes all the mass and is relatively compact, whereas in CA particles, 
  monomers form separate clusters which then aggregate, thus resulting in a smaller
  fractal dimension of the aggregate.
  The PA process occurs when the 
  number of monomers compared to the available volume is high, 
  increasing the chances of collision amongst small aggregates. 
  
  Depending on the physical conditions of the primordial protosolar nebula, 
  in terms of dust to gas ratio and dust composition, we can expect each of these kinds
  of aggregates to be formed \citep{weidenschilling_origin_1997, kimura_light-scattering_2001}. 
  They have also each been produced by computer simulations
  and laboratory experiments simulating the initial stages of planetary accretion 
  \citep{meakin_fractal_1991, blum_growth_2008}. 

\subsection{Flattening simulation}

\begin{figure}
	\center
	\includegraphics[width = 8cm]{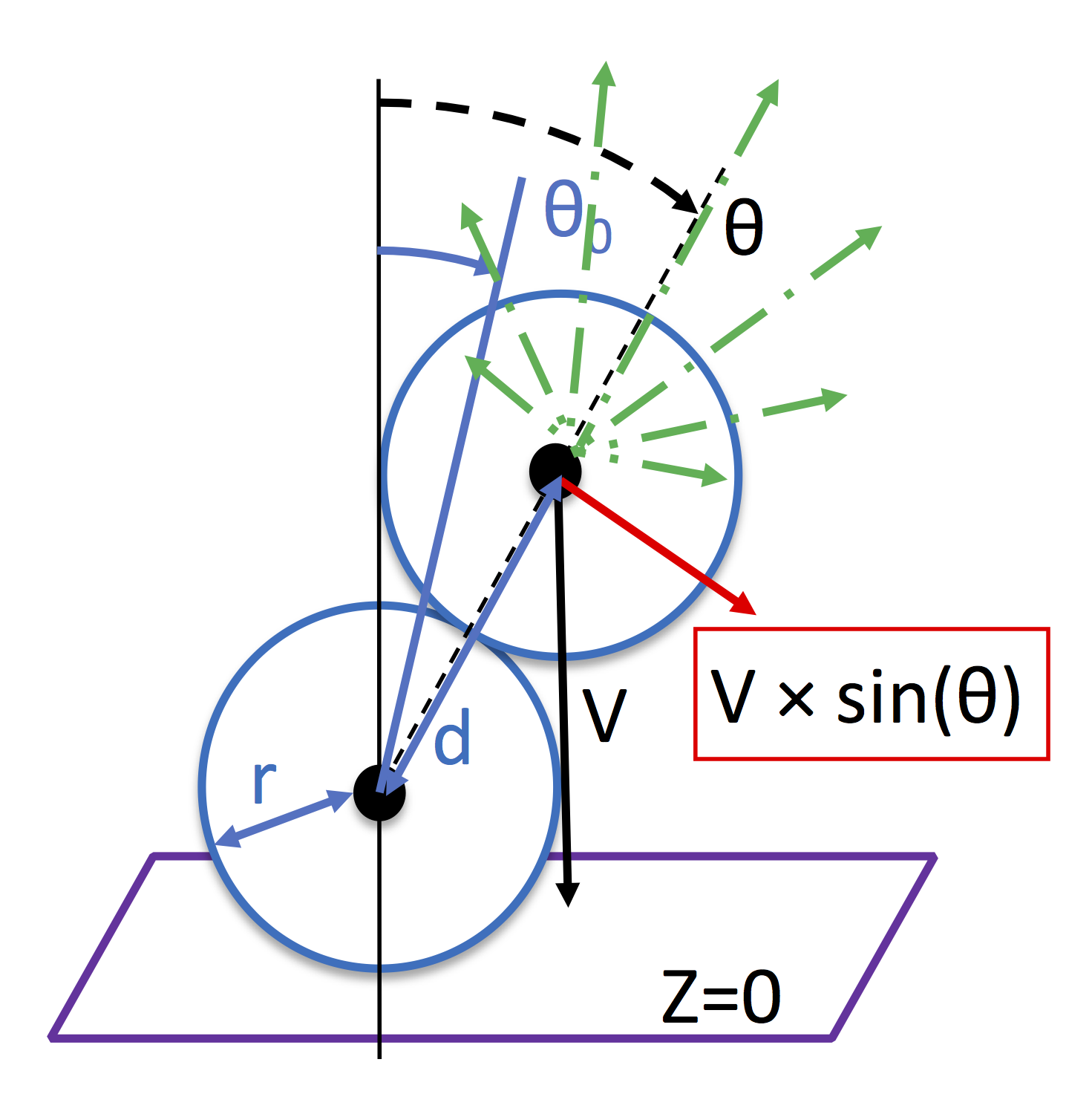}
	\caption{Representation of the collision geometry for 2 superposed monomers. 
	If $\theta < \theta_0$, then the superposition remains stable.
	When $\theta > \theta_0$ the cohesive link between the monomers is broken 
	and the upper monomer will bounce randomly following some of the 
	green arrows and will attach itself to the $z=0$ surface at a further point, 
	thus fragmenting the aggregate.}
	\label{fig:collision}
\end{figure}

  In a first step, 3D off-lattice aggregates of a number 
  $N$=10 000 identical spherical particles (called monomers) 
  are generated according to the 4 different aggregation processes 
  described above. The resulting fractal aggregates are characterized by different 
  initial fractal dimensions according to the approximate relationships: 
  DLCA ($D_f \approx 1.8$), RLCA ($D_f \approx 2.1$), 
  DLPA ($D_f \approx 2.5$) and RLPA ($D_f \approx 3$). 
  The values were calculated using the well known self-similarity properties of 
  fractals whereby the number, $N_m$, of monomers constituting the aggregate located 
  within a sphere of radius $R$ follows $N_m\propto R^{D_f}$, where $R$ 
  is smaller than the gyration radius of the aggregate.
  The gyration radius of a fractal aggregate is a measure of the 
  extent of the aggregate, akin to the standard deviation of the monomers' 
  distance to the centre of mass of the aggregate and can be calculated by 
  \begin{equation} 
  R_{g}^2=\frac{1}{2N}\times \sum_{i,j} (r_i-r_j)^2 = \frac{1}{N}\times \sum_{i} (r_i-r_c)^2
  \end{equation}
  where $N$ is the number of monomers in the aggregate,  $r_i$ and $r_j$ 
  are the spatial coordinates of the center of the monomers $i$ and $j$, 
  and $r_c$ corresponds to the spatial coordinates of the center of mass of the 
  aggregate \citep{jullien_aggregation_1987}.
  A representation of each of the four aggregate types is given in Fig.~\ref{fig:aggregates}.
  These aggregates may correspond to different types of cometary particles as ejected 
  from the surface of the nucleus by gas pressure. 
  For each aggregate type, 1000 different aggregation simulations 
  were performed to statistically analyze the results. 

\begin{figure*}[ht]
  \begin{adjustbox}{addcode={\begin{minipage}{\width}}{\caption{%
      Figure summarising the effect of initial particle morphology ($D_f$) 
      and bond cohesion on the morphology of the flattened particles.
      A scale is given in number of monomers, and a referential frame 
         is indicated by colored arrows (x=blue, y=red, z=green).
      	\label{fig:tab_agg}}\end{minipage}},rotate=90,center}
      \includegraphics[scale=.65]{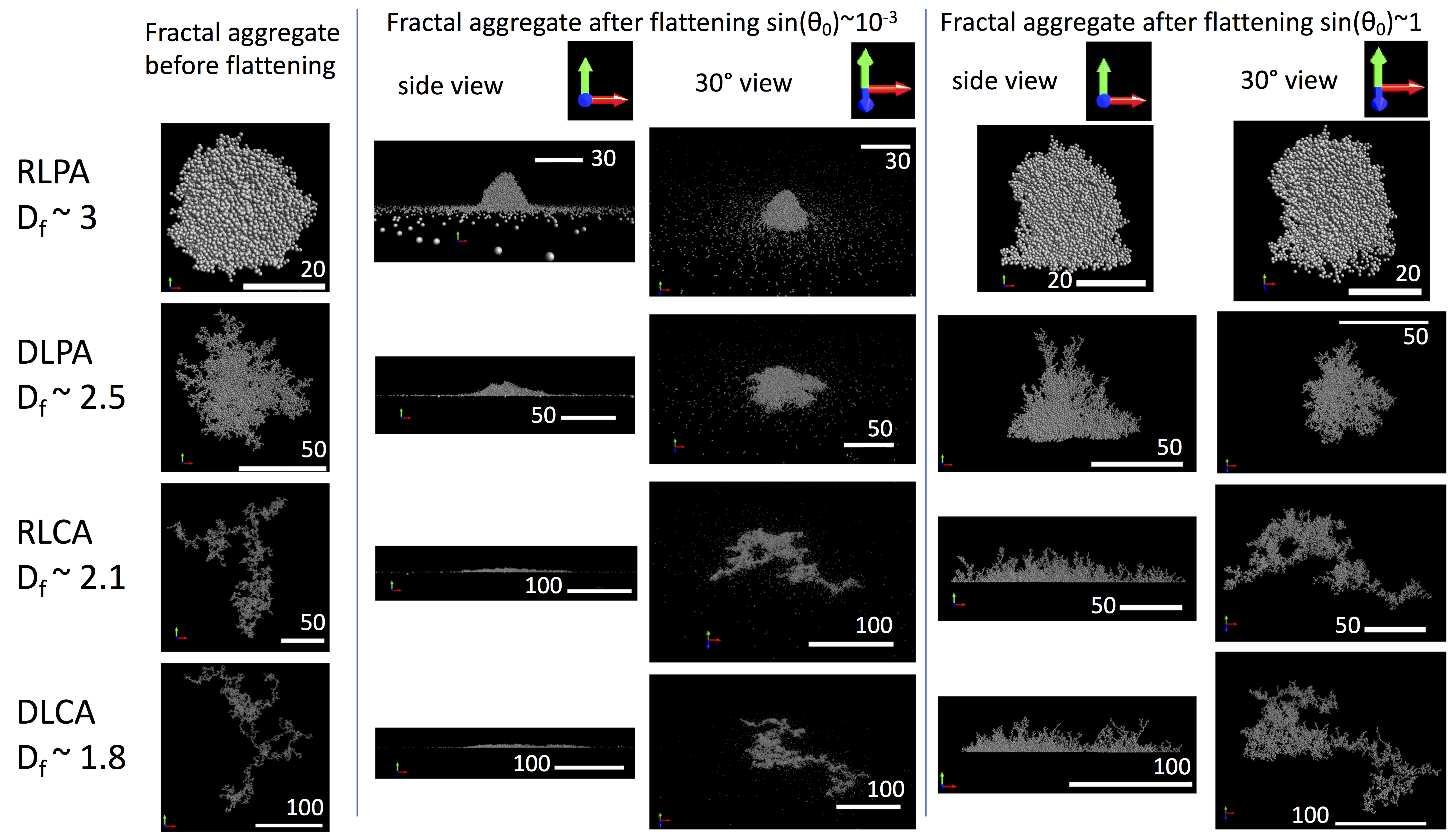}%
  \end{adjustbox}
\end{figure*}

  In a second step, simulating the particle collection and flattening 
  observed by COSIMA during the Rosetta mission, 
  the aggregates are projected monomer by monomer onto 
  the plane $z = 0$ as shown in Fig.~\ref{fig:collision}. 
  The monomers are selected iteratively by increasing $z$ values. 
  If a monomer is projected directly onto the plane  $z = 0$ without 
  encountering any other monomer, it sticks directly to the collision plane. 
  In the case where it encounters a monomer that is previously stuck under it, 
  we consider that a bond exists between the two monomers
  and that it  will be broken 
  if $E_{VdW}<E_{k}$ where $E_{VdW}$ is the van der Waals energy and $E_{k}$ is 
  the kinetic energy of the incoming particle. This condition can be written as in 
  Eq~(\ref{eq:collision}) considering van der Waals interactions between 
  the two spherical elements.
\begin{equation} 
	E_{VdW}=\frac{A_H r}{12d} \leq E_{k}=\frac{2}{3}\pi r^3 \rho (\sin{\theta})^2 V^2
	\label{eq:collision}
\end{equation}
\begin{equation} 
	(\sin{\theta_{0}})^2 = \frac {A_H} {8 \pi r^2 \rho d V^2} 
	\label{eq:Vsin}
\end{equation}
  where $E_{VdW}$ is the van der Waals energy, $A_H$ is the Hamaker constant 
  for the material considered, $r$ is the radius of a single monomer, $d$ is the 
  diameter of a monomer, $E_k$ is the kinetic energy of the monomer, $\rho$ is the 
  density of the monomer, $\theta$ is the angle between the direction 
  linking the two centers of the monomers and the vertical direction, as illustrated 
  in Fig.~\ref{fig:collision} and $V$ is the velocity of the aggregate with respect 
  to the collecting surface $z=0$.9
  The monomer diameter, $d$, is slightly larger 
  (by \SI{0.4}{\nano\meter}) than the steady state 
  distance between the centers of two touching monomers. 
  Typical monomer diameters are considered to be \SI{20}{\nano\meter} or larger, 
  making this difference negligible ($< 2\%$). We thus consider the distance between two 
  touching monomer centers to be equal to $d$.
  The Hamaker constant of two particles interacting corresponds to a measure of the 
  relative strength of the particles material with respect to the attractive van der Waals 
  forces between them  \citep{hamaker_londonvan_1937}.

  We call $\theta_{0}$ the angle $\theta$ for which Eq~(\ref{eq:collision}) is an equality. 
  $\sin{\theta_{0}}$ is a threshold above which monomers may break their bonds and bounce. 
  Changing this parameter can either be viewed as changing the cohesive strength between monomers 
  or as changing the collection velocity, as Eq~(\ref{eq:Vsin}) shows. 
  Thus, with the Hamaker constant of dry minerals under vacuum 
  conditions $A_H \approx$\SI{1e-19}{\joule} \citep{israelachvili_intermolecular_2011}:
\begin{itemize}
	\item $\sin{\theta_0} \approx 1$ corresponds to very cohesive monomer bonds, 
	low collection speed or very small monomer size (value typically obtained for 
	$V =$ \SI{1}{\meter\per\second} and  $r =$\SI{0.01}{\micro\meter} or for 
	 $A_H$ values higher than \SI{1e-19}{\joule})
	\item $\sin{\theta_0} \approx 10^{-3} $ corresponds to all bonds being broken, 
	relatively higher collection speed, or larger monomer sizes 
	(value typically obtained for $V =$\SI{10}{\meter\per\second} and  $r =$\SI{0.1}{\micro\meter})
\end{itemize}
  So, if a projected monomer meets another monomer, we can compute  a collision parameter 
  ($\nu = \frac{\sin {\theta}}{\sin {\theta_0}}$). If $\nu \leq 1$, the monomer sticks to the one it bumps into. 
  If $\nu > 1$ , the incoming monomer bounces according to a random direction based on 
  the Lambertian reflection rule (see Fig.~\ref{fig:collision}) and sticks to 
  the plane $z = 0$ or previously stuck monomers if they are present.

  To make the model more realistic with respect to potential mass loss that may be incurred 
  by the aggregates as they are flattened and their bonds are broken, in further simulations 
  a mass loss probability $P_{loss}$ is introduced. 
  In this case, if a monomer meets the condition $\nu > 1$, then it will be removed 
  from the simulation with the mass loss probability $P_{loss}$ 
  which matches the chance that some monomers 
  do not stick to any others and bounce back to free space during the collision. 

  An illustration of the effect of changing the value for $sin(\theta_{0})$ is given 
  in Fig.~\ref{fig:tab_agg} where the morphology of flattened aggregates is clearly 
  dependent upon the initial structure of the aggregates and the geometric parameters. 
  Under conditions where most bonds are broken ($\sin{\theta_0} \approx 10^{-3} $), 
  the more compact aggregates appear to generate 
  a small pyramid of monomers with an angle of repose. 
  The more porous the aggregates, the flatter appears to be the projection. 
  In the case where bonds are unbroken ($\sin{\theta_0} \approx 1$), 
  similar structures appear, but some vertical chain-like 
  columns of monomers extending upwards are also present and 
  increase the relative height of the flattened aggregate. 
  These columns of monomers appear due to the increased strength 
  of the bonds between the monomers, forming chain-like vertical structures 
  that are not broken by the flattening geometry (as $\theta_0 > \theta$ over the monomers' column).
  We therefore see that both parameters ($D_f$ and $\theta_{0}$) influence significantly 
  the outcome of the simulated projection. 

\begin{figure}
	\center
	\includegraphics[width = \columnwidth]{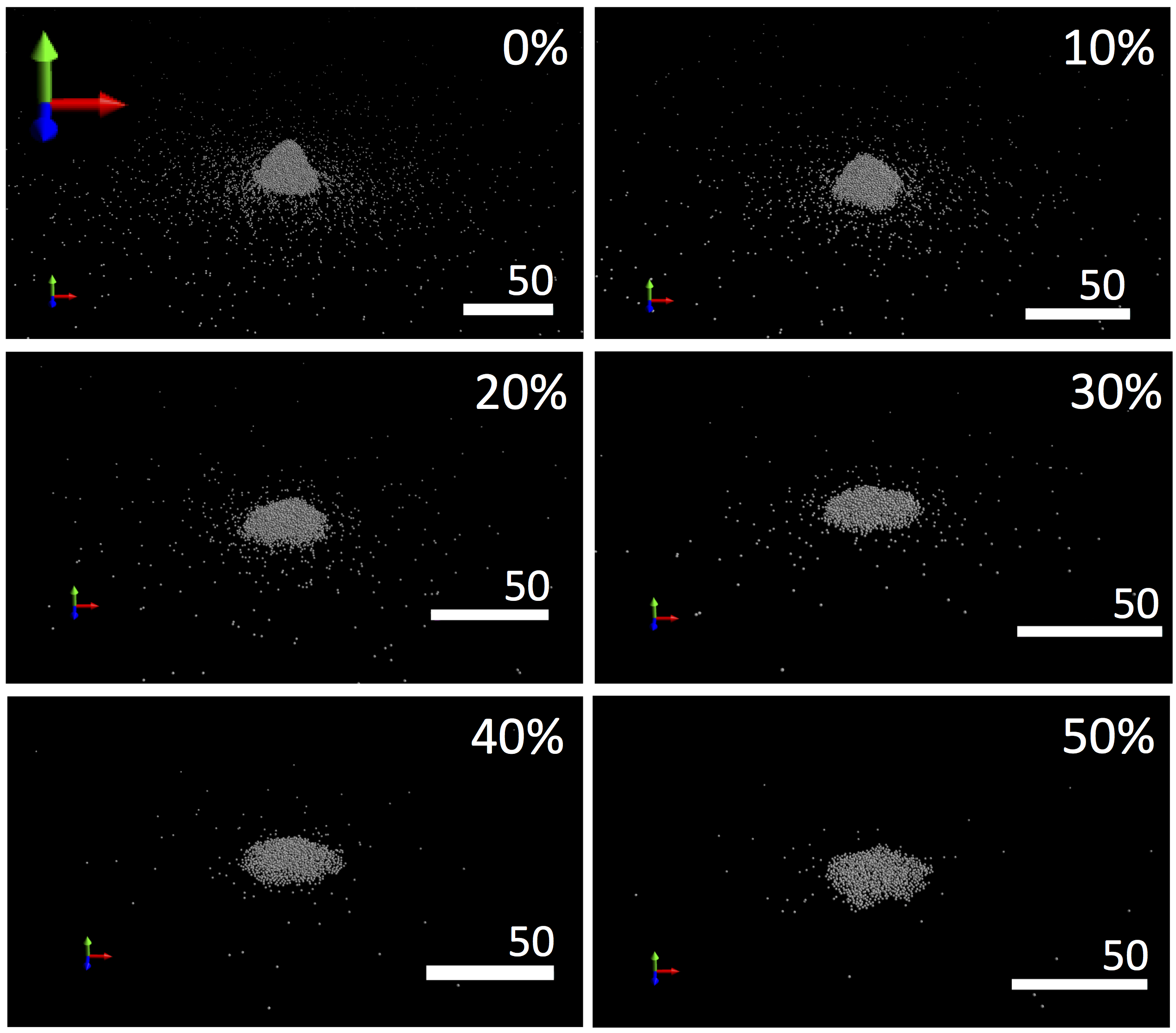}
	\caption{3D view of a RLPA aggregate after projection with different mass loss.
	A scale is given in number of monomers, and a referential frame 
         is indicated by colored arrows (x=blue, y=red, z=green) with an approximate 
         \SI{30}{\degree} viewing angle.}
	\label{fig:massloss}
\end{figure}

  Fig.~\ref{fig:massloss} shows the resulting projections in the case where the monomers 
  have a non-zero probability to bounce back to space due to mass loss processes. 
  In this case we only consider RLPA aggregates with different $P_{loss}$ values 
  ranging from 0\% to 50\%.
  As the mass loss probability gets larger, only a flat footprint of the aggregate 
  remains with a very low aspect ratio which can represent the results of the low speed laboratory 
  aggregate sticking experiments of \citet{ellerbroek_footprint_2017} where most of 
  the initial aggregate mass was lost.
  Such mass loss processes may also be at work during the COSIMA particle 
  collection.

\section{Results}

\subsection{Data Analysis}

\begin{figure}
	\center
	\includegraphics[width = \columnwidth]{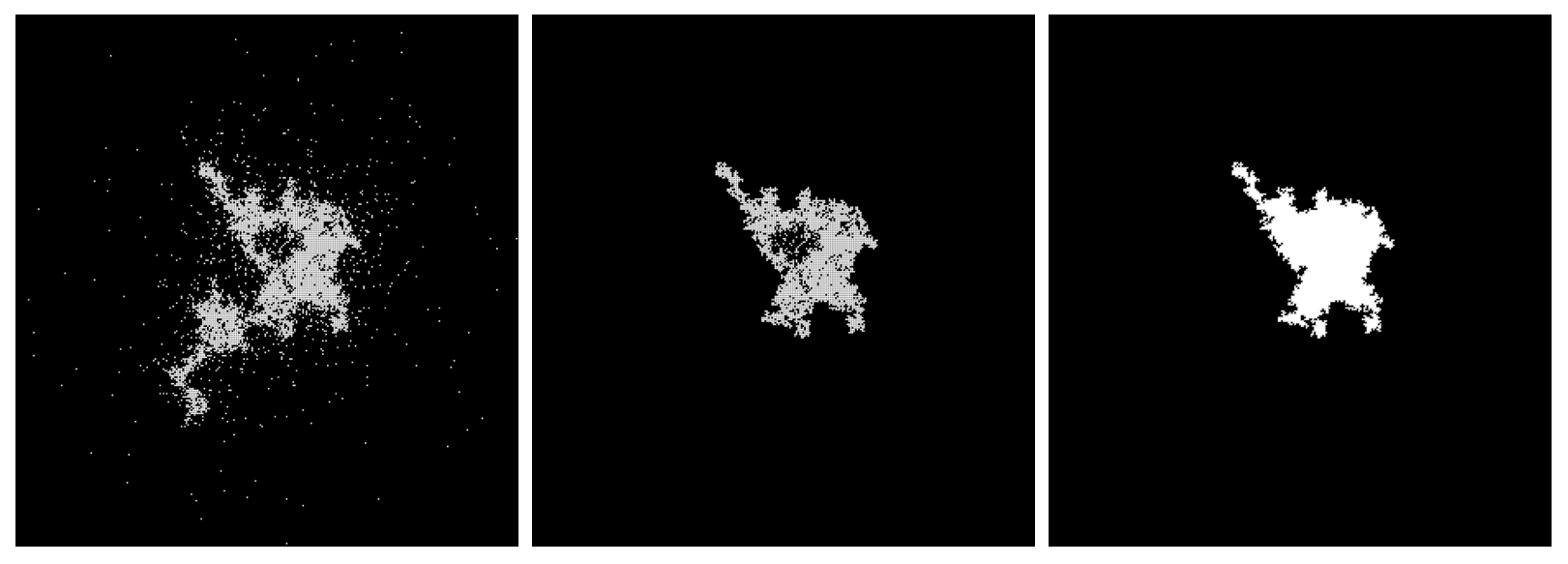}
	\caption{Illustration of the connected area calculated for 
	a flattened aggregate of type RLCA. The flattened particles seen from above 
	is shown on the left. The calculated connected areas are shown with gaps 
	in the middle and without gaps on the right. The parts of the aggregate
	that are not connected to the largest connected aggregate are removed from 
	the processing.}
	\label{fig:area_computation}
\end{figure}

  The aggregate flattening simulations were run to create 1000 aggregates of 
  each of the four types, using 10,000 monomers each, for the four fractal dimensions
  considered, with the $\sin(\theta_{0})$ parameter ranging from 1 to $10^{-6}$ and
  with a $P_{loss}$ probability of mass loss ranging from 0 to 0.5. 
  This was done in order to obtain good statistics for the aspect ratio 
  of each numerically flattened  aggregate for comparison to the COSIMA measurements. 
  The height, $H$, of the flattened aggregates is the maximum 
  value of $z$ among all the sticking monomers. 
  To compute the area, $A$, we considered only 
  the monomers visible from above (looking towards the $-z$ direction) and, 
  based on their position, we calculated the 
  contour of the projected connected set of monomers \citep{lorensen_marching_1987}. 
  We computed two different connected areas: one with gaps and one without gaps as 
  Fig.~\ref{fig:area_computation} shows. The area with gaps is always somewhat 
  smaller than the area without gaps but is essentially linearly correlated with it. 
  Therefore, we calculated the results based on the connected area without gaps. 
  In this way, we can calculate a statistical distribution of 
  the aspect ratio, $H / \sqrt{A}$, for particles 
  of each kind similar to the procedure used with the COSIMA data
  and assess the effect of the different parameters on the morphology of 
  the flattened aggregates. 

\subsection{The morphologies of flattened aggregates}

\begin{figure}
	\center
	\includegraphics[width = 2.8in]{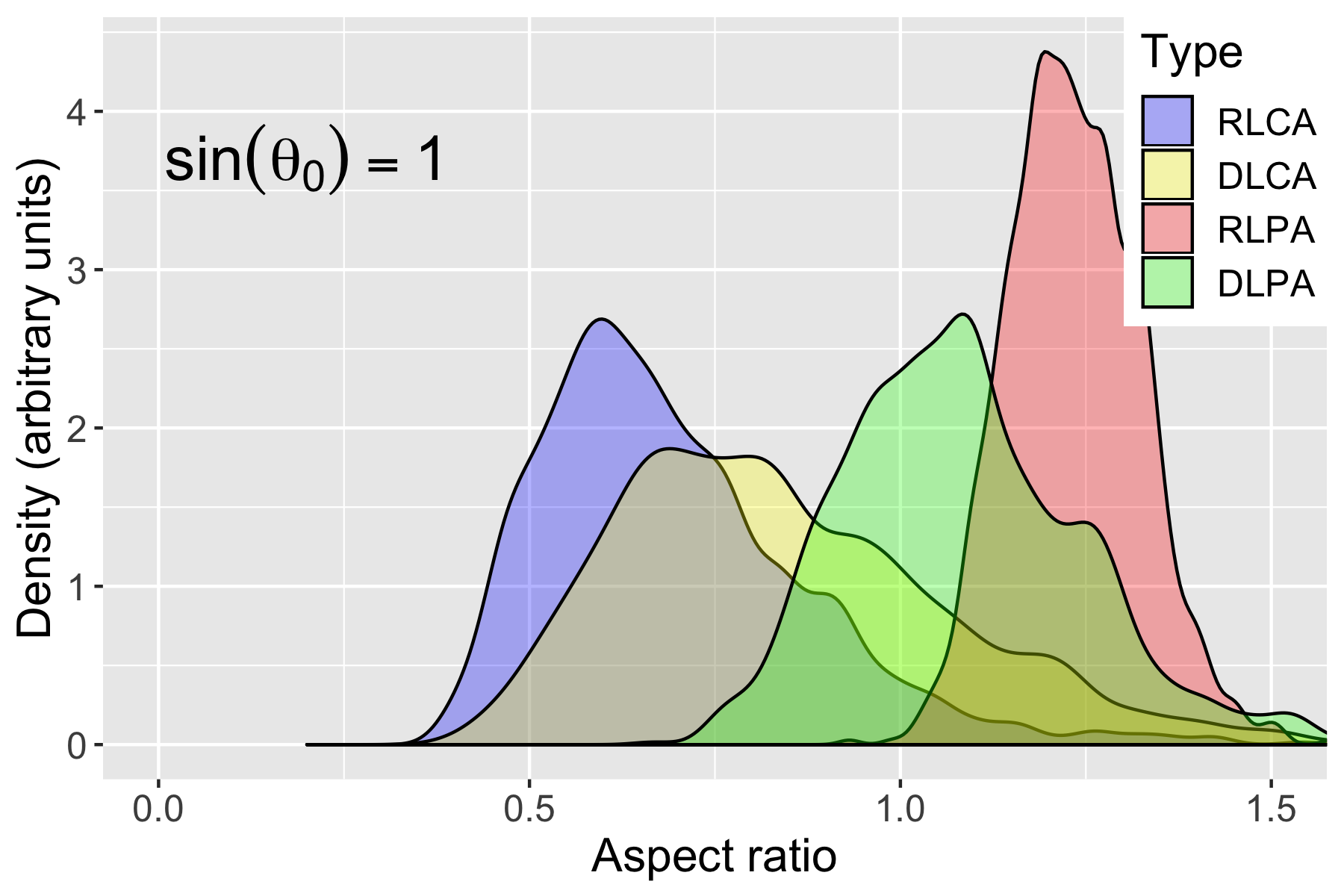}
	\includegraphics[width = 2.8in]{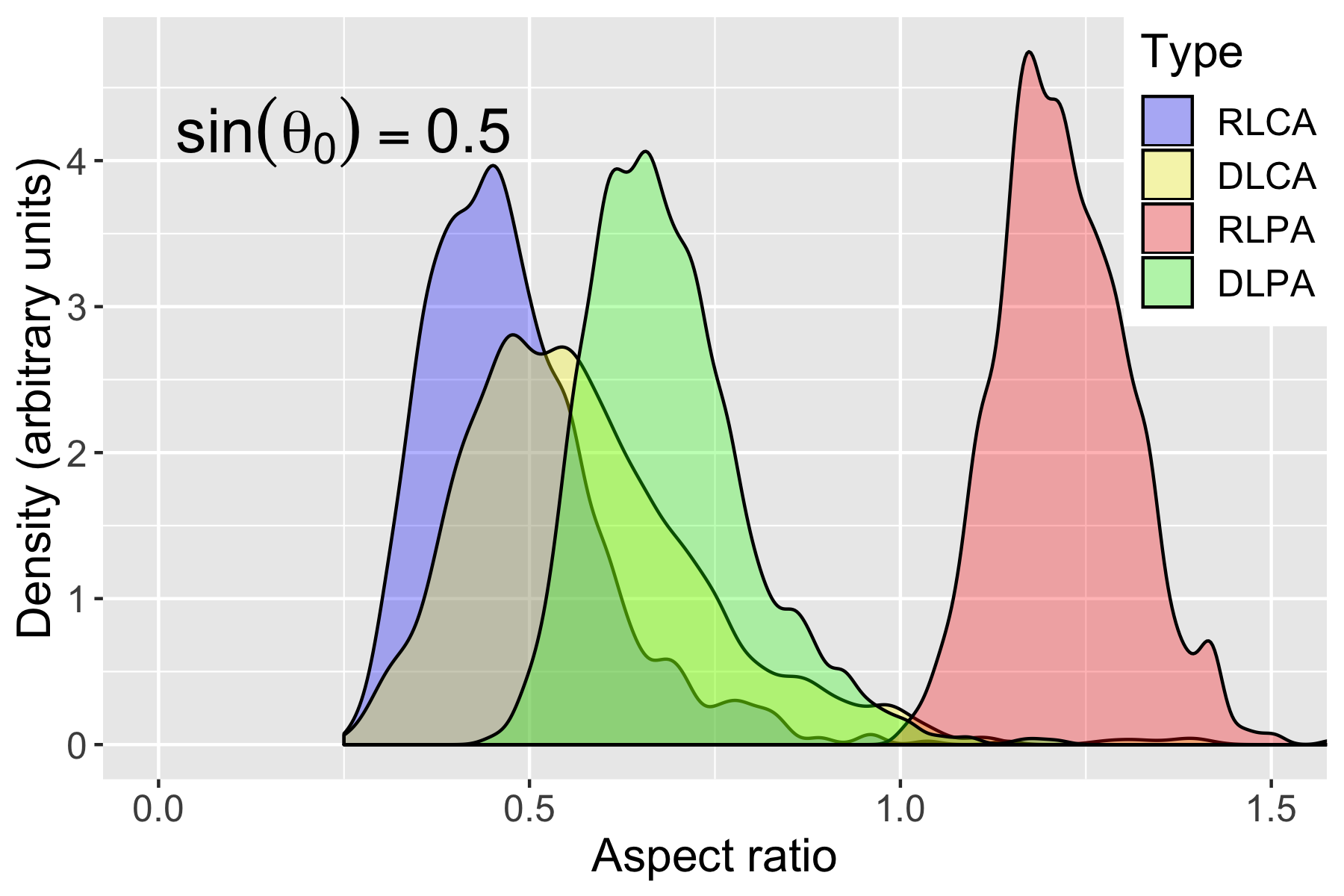}
	\includegraphics[width = 2.8in]{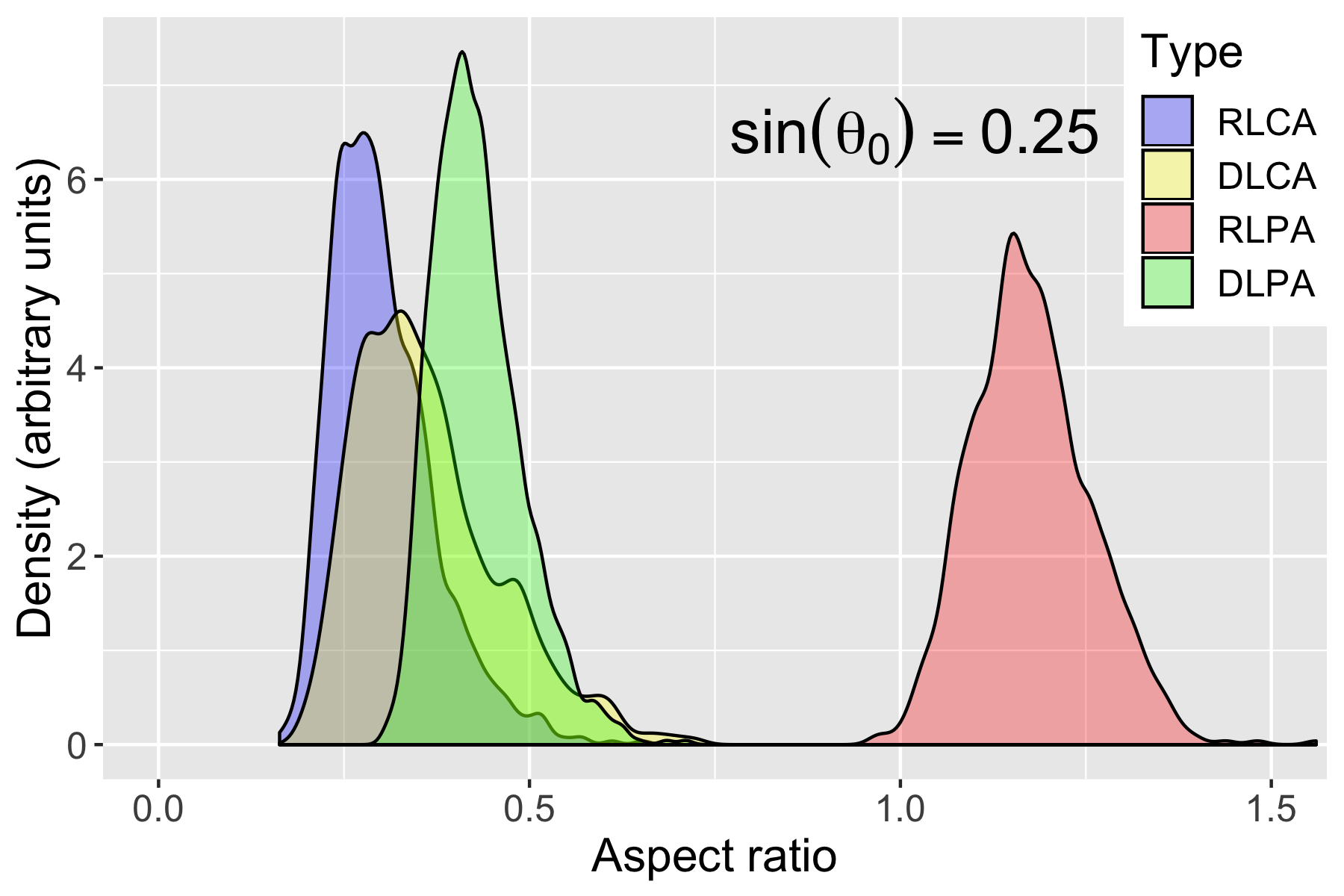}
	\includegraphics[width = 2.8in]{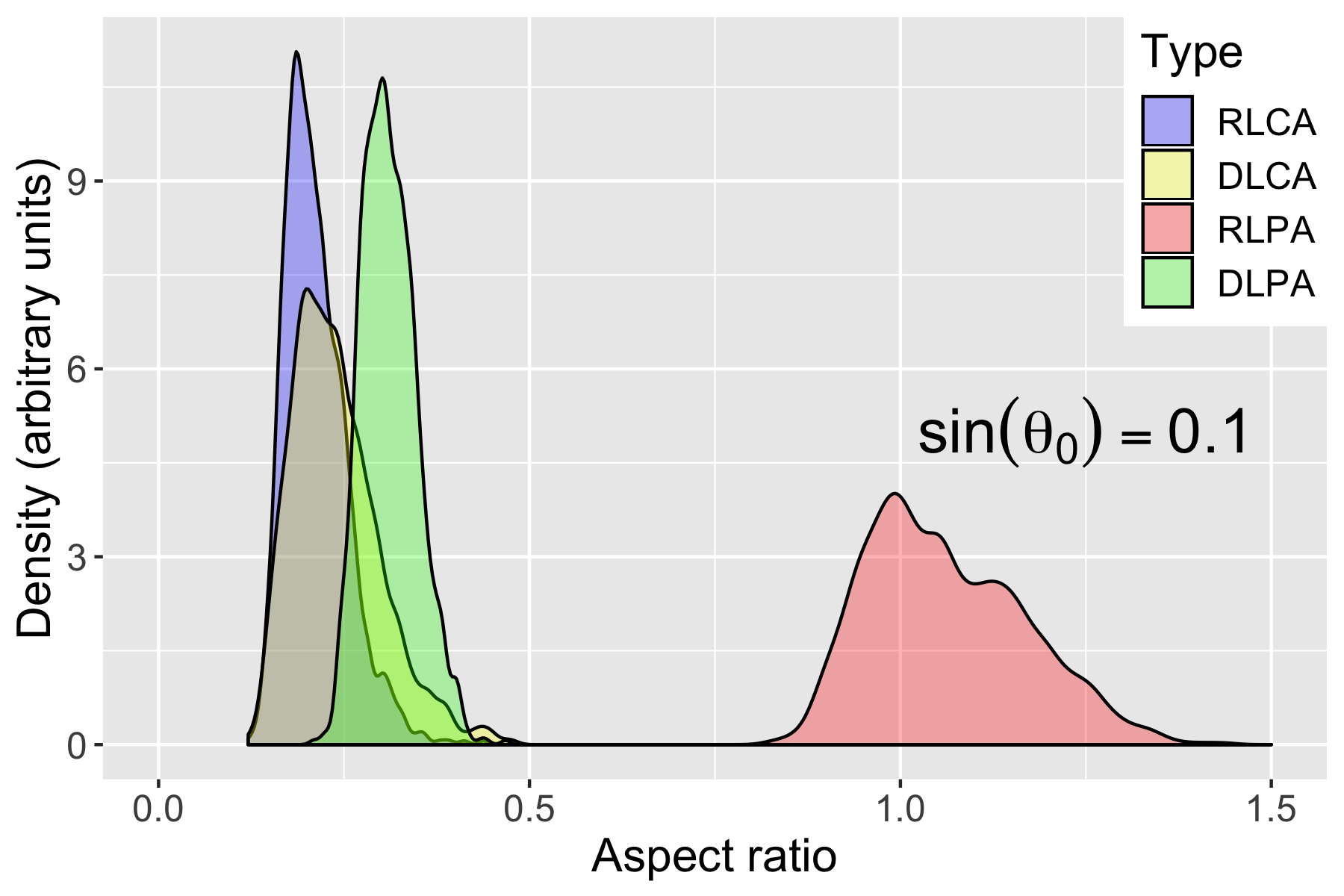}
	\includegraphics[width = 2.8in]{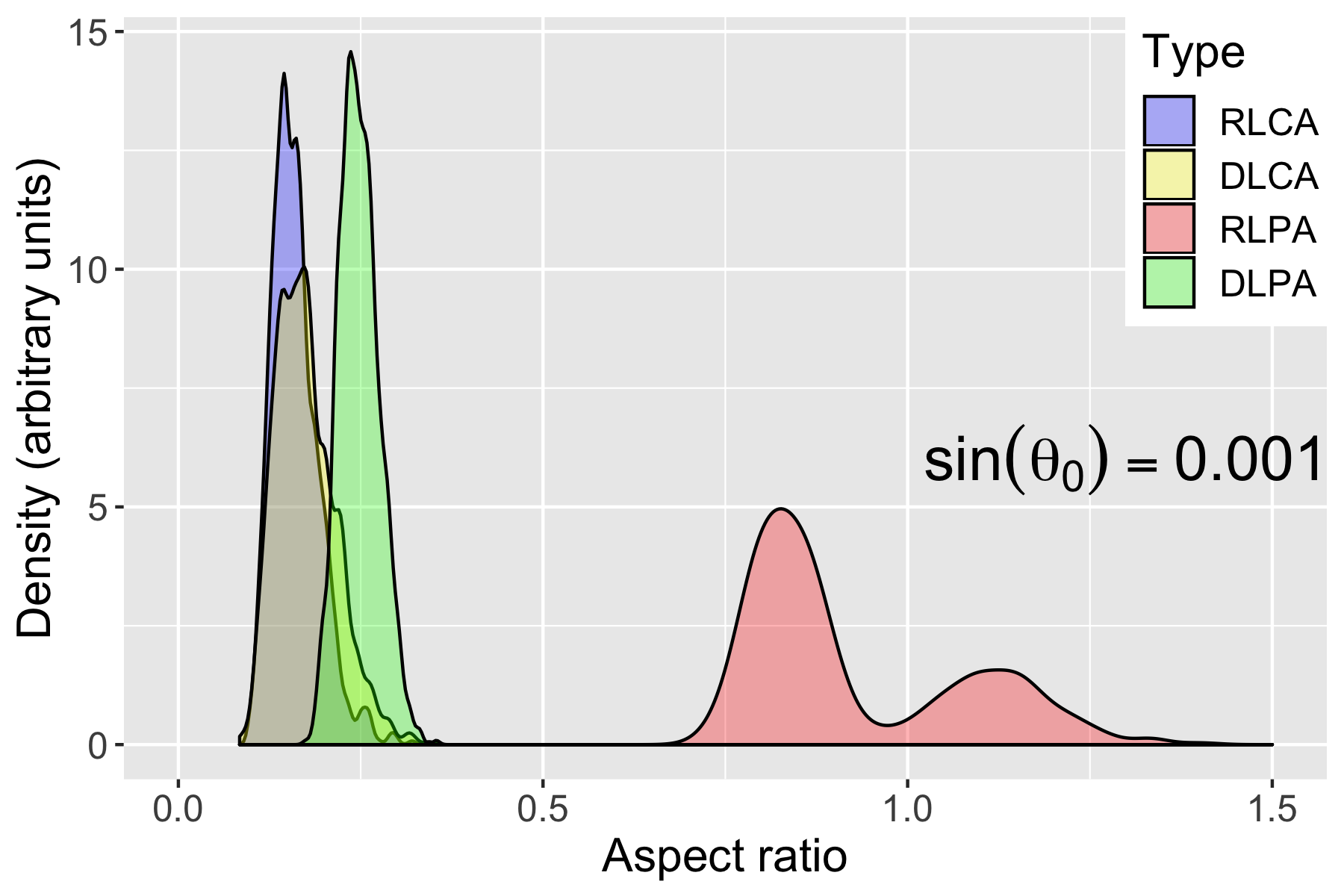}
	\caption{Distribution of aspect ratio, $H/\sqrt{A}$, for 1000 aggregates of each type 
	(RLCA, DLCA, RLPA, DLPA) with $\sin(\theta_0)$ ranging from 1 to 0.001 without mass loss.}
	\label{fig:effect_sintheta}
\end{figure}

  Figure~\ref{fig:effect_sintheta} represents the density distribution of aspect ratios 
  calculated for the 1000 flattened aggregates of each fractal type and for 4 different values 
  of $\sin(\theta_0)$. 
  The upper figure is calculated for a $\sin(\theta_0)=1$ corresponding to a simulation
  where no bond between monomers is broken (illustrated on the right hand side of the 
  Figure~\ref{fig:tab_agg}). One can see that the distribution of aspect ratios overlaps 
  between about 0.5 (relatively flat aggregates) and 1.3.
  The distribution also separates relatively well the different types of aggregates 
  with the more compact aggregates of type PA having a median aspect ratio value
  of 1.18 ($\sigma=0.15$) and the fluffier aggregates of type CA having a
  median aspect ratio value of 0.73 ($\sigma=0.22$).
  Therefore, to first order, the process appears to separate the aggregates with 
  fractal dimensions above or below 2 into two groups. 
  This is somewhat expected since more compact aggregates will present more 
  opportunities for solid vertical structures of monomers to remain unbroken and 
  to vertically extend the projected aggregate. 
  One can also notice that the aspect ratio distributions of the CA type aggregates 
  present an extended right wing showing that some of those aggregates could 
  still have aspect ratios close to one, if their monomer bonds are strong compared
  to the energy of impact.  
  
  As $\sin(\theta_0)$ decreases, the number of broken bonds increases and the 
  projected aggregates get flatter. The minimum aspect ratio decreases and 
  reaches 0.1 for values of $\sin(\theta_0)=0.1$ or lower. The distribution of the most
  compact particles (RLPA with $D_f \approx 3$) is now clearly separated from the 
  distribution of the other aggregates and remains around 1, indicating that the 
  surface dimensions covered by the flattened aggregate in $x$ and $y$
  are of the same order of magnitude as its vertical extent in $z$.
  With respect to the distributions of the less compact aggregates, 
  we notice that the distributions for CA aggregates with fractal dimensions 
  lower than about 2 become quickly undistinguishable. Those flattened aggregates
  would therefore present essentially the same aspect ratio distributions irrespective 
  of their initial morphology. The DLPA aggregates that have a fractal dimension around 2.5 
  are located in between those two extremes and clearly separated from them at low
  values of $\sin(\theta_0)$. For example, the standard deviation of the distributions for 
  $\sin(\theta_0) = 0.25$ range from 0.06 to 0.09. 
  The DLPA distribution average aspect ratio is approximately 0.3 for $\sin(\theta_0) =0.1$
  or lower.
  At values of  $\sin(\theta_0)$ lower than 0.1, the density distributions stabilize towards
  their final values. 
  One can also notice a bimodal density distribution for the 
  flattened RLPA aggregates, corresponding to whether vertical columns of monomers
  appear within the pyramid somewhat extending its height.
  We expect the random size distributions of monomers in real dust aggregates 
  to limit the aspect ratio to the lower values of around 0.75-1.0. 
  Some similar linear chain-like structures were also detected in the analysis of 
  COSIMA particles, such as the 2CF Adeline particle \citep{hornung_first_2016}.
 
\begin{figure}
	\center
	\includegraphics[width = \columnwidth]{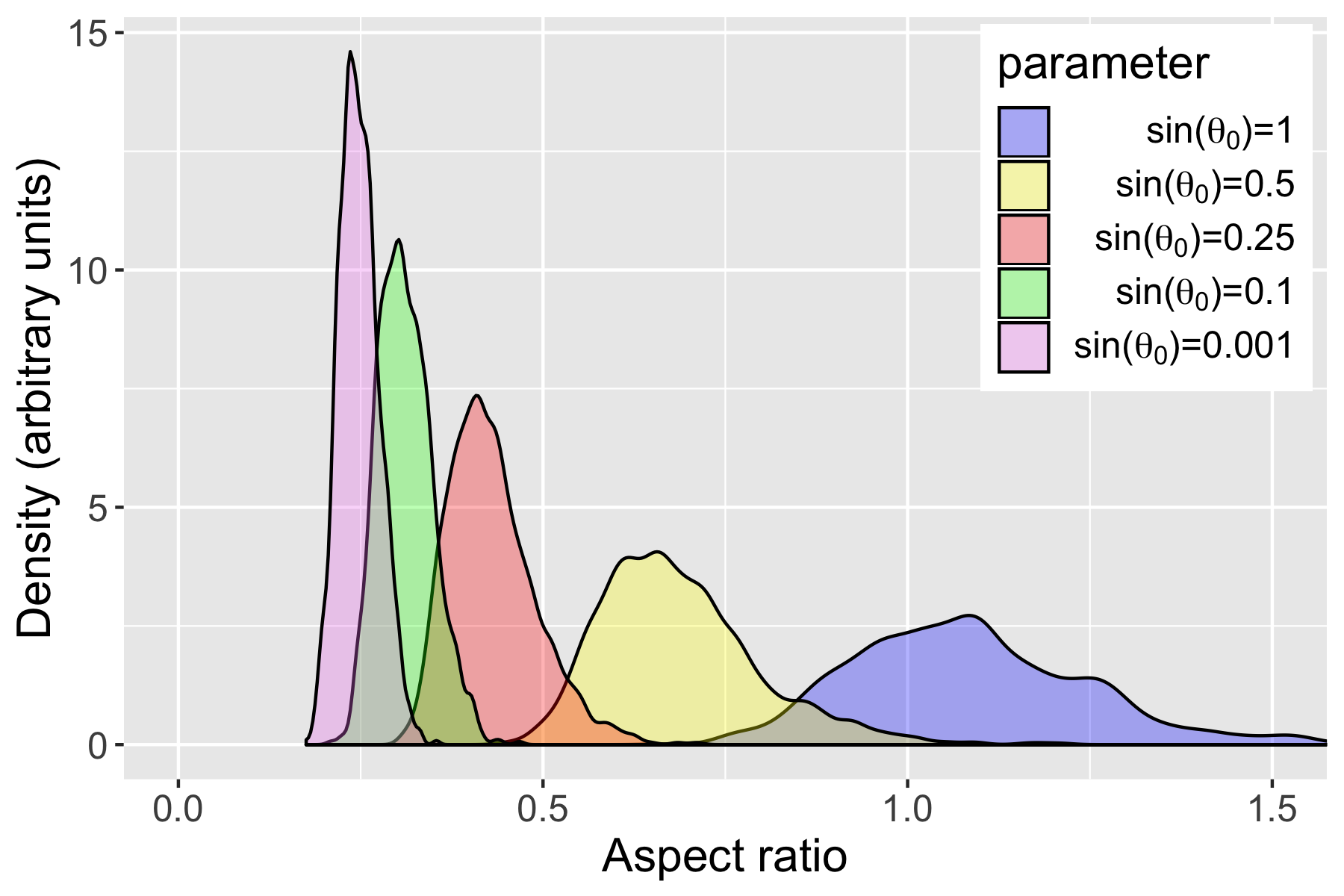}
	\caption{Distribution of aspect ratio, $H/\sqrt{A}$, for DLPA aggregates with different 
	$sin(\theta_0)$ without mass loss.}
	\label{fig:effect_Df}
\end{figure}
 
  The effect of the $\sin(\theta_0)$ parameter is further illustrated in Figure~\ref{fig:effect_Df}
  where aspect ratio distributions of DLPA aggregates are calculated for different values of 
  $\sin(\theta_0)$ ranging from 0.001 to 1 and are superposed. 
  As the $\sin(\theta_0)$ value decreases, the aspect ratio decreases (due to the larger 
  number of bonds breaking) from approximately 1 to  0.25. 
  One  can also notice that the standard deviation of the density distribution
  also decreases, indicating that most aggregates of this type flatten in the same way. 
  This is related to the randomization of monomer deposition after bond breaking
  which reduces the range of vertical extent possible after flattening. 
  Based on this figure, we can see that given a relatively narrow range 
  of collection velocities, since equation~\ref{eq:Vsin} indicates that $\sin(\theta_0)$ 
  is proportional to $\frac{1}{V}$,
  a large range of bond cohesive strengths in the aggregates would lead to 
  a larger range of aspect ratios for the same initial structure of the aggregate. 
  This is especially true of DLPA as the aspect ratios for these particles range from 0.25 to 1.2.
  Compact aggregate aspect ratios would range between 0.8 and 1.3, while  
  aggregates with fractal dimensions around 2 and lower  present aspect ratios 
  ranging from 0.1 to 1. 
  If the cohesive strength of monomer bonds in the aggregate are randomly distributed
  one can expect to detect more aggregates with small flattened aspect ratios than large 
  flattened aspect ratios. 
  
\begin{figure}
	\center
	\includegraphics[width = \columnwidth]{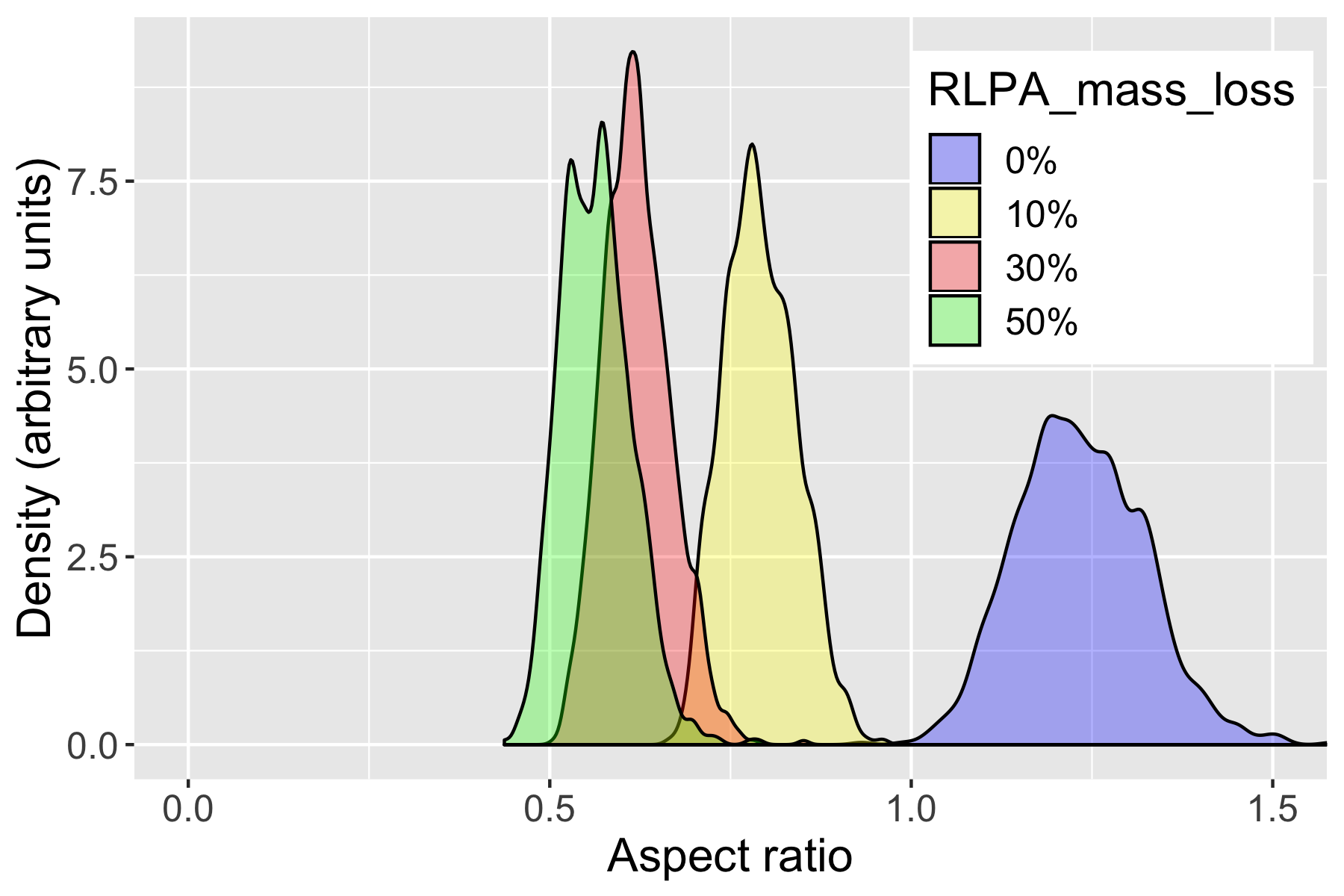}
	\includegraphics[width = \columnwidth]{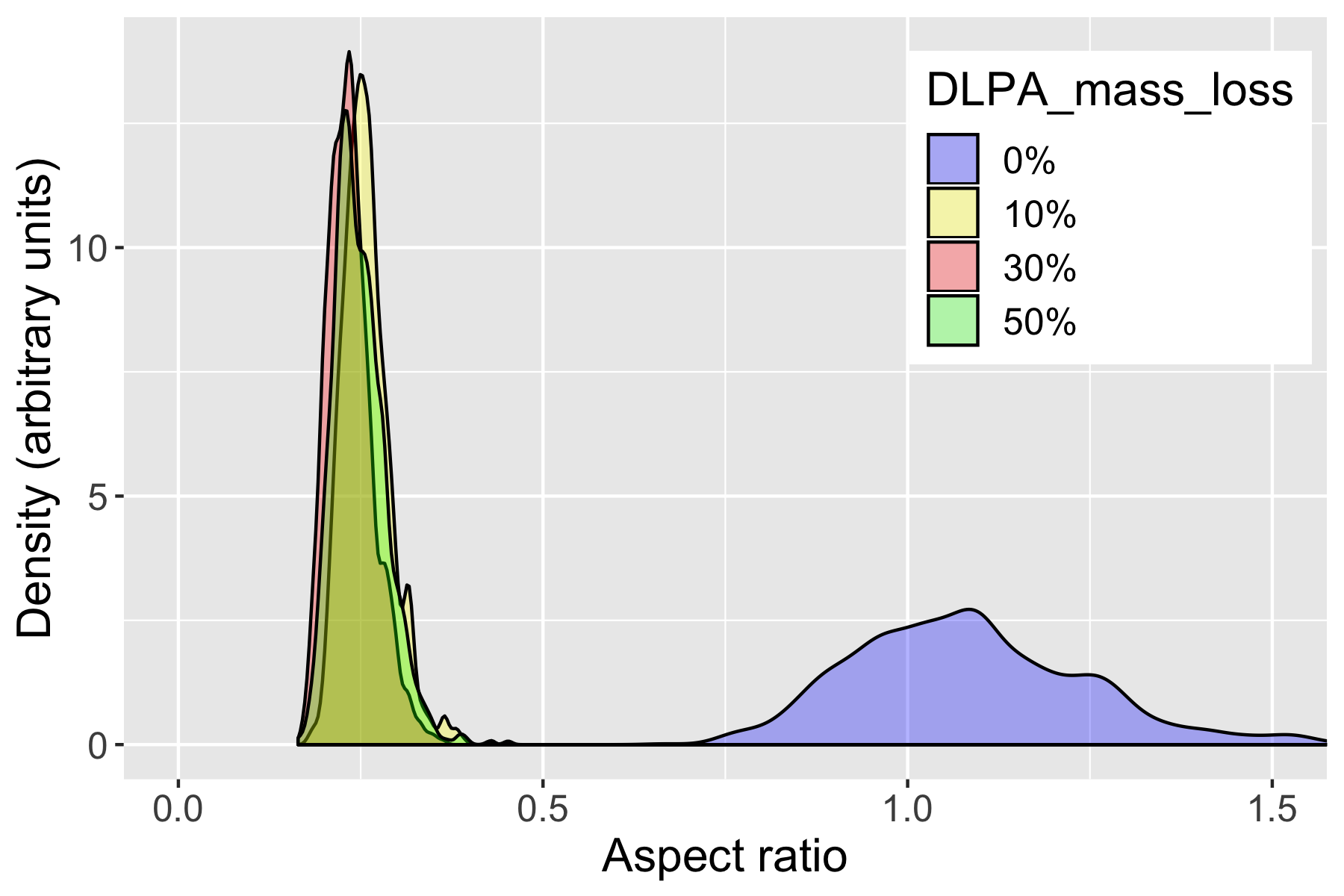}
	\caption{Distribution of aspect ratio, $H / \sqrt{A}$, for RLPA aggregates (top)
	and DLPA aggregates  (bottom)
	with $sin(\theta_0)=1$  and different mass loss probability coefficients.}
	\label{fig:effect_massloss}
\end{figure}
  
  Finally, the effect of the mass loss coefficient is illustrated in Figure~\ref{fig:effect_massloss} top
  where the probability density of aspect ratios for RLPA aggregates with $sin(\theta_0)=1$ is calculated
  for mass loss parameters ranging from 0\% to 50\%. 
  As expected, the mass loss parameter reduces the aspect ratio of the flattened aggregates 
  because of the loss of monomers. As compared to the variation in aspect ratio 
  distribution for varying $\sin(\theta_0)$, one can notice that the end aspect ratio distribution 
  remains relatively large (larger than 0.5) which is due to the simultaneous loss of monomers 
  in all directions, so that the dimensions of the flattened aggregate are reduced
  in all dimensions at more or less the same rate (in $x$, $y$, and $z$).  
  This parameter is also important in reducing the final aspect ratio of the flattened aggregates. 
  
  In the case of DLPA aggregates and the more fluffy ones, the initial aggregate is so porous
  that even moderate mass loss destroys the structure during flattening. 
  This leads rapidly to a very flat final projected structure as illustrated in Figure~\ref{fig:effect_massloss} 
  bottom.
  
  It therefore appears that the initial fractal dimension of aggregates strongly affects the morphology 
  of flattened aggregates, and that, depending on the effect of parameters such as 
  the speed of collection, strength of bonds between the monomers and mass loss fraction,
  it may, or may not, be possible to distinguish the initial structure of the particle from their
  flattened morphologies. 

\section{Discussion}
  
\subsection{Comparison with COSIMA observations}

  The aspect ratio variation with initial $D_f$ (aggregate type), 
  $sin(\theta_{0})$ and $P_{loss}$ can be compared with the values 
  observed by COSISCOPE and presented in Fig.~\ref{fig:langevin}. 
  On the one hand, only the PA aggregate types have an aspect ratio large 
  enough to explain the presence of the compact particles in the COSIMA 
  aspect ratio distribution. This implies that a population of particles
  with fractal dimension between 2.5 and 3 must be present in the distribution 
  of particles ejected by 67P.
  
  On the other hand, in order to explain the presence of morphologies with 
  aspect ratios as low as 0.1 to 0.3, where the distributions of COSIMA
  particles of type G, R and S peak, other types of particles or processes need to be invoked. 
  From our simulations, even with a mass loss as large as 50\%, 
  RLPA aggregates alone cannot explain the range of aspect ratio observed. 
  However, the DLPA type particles could reach aspect ratio values as low as 0.2 
  either with different cohesive strengths and/or velocities 
  ($\sin(\theta_0)$) or with mass losses up to 50\%. 
  Finally, a fractal dimension lower than 2 would also lead to very low final aspect ratios
  even when considering particles with higher cohesive strengths. 
  The large range of distribution observed by COSIMA could therefore be explained by: 
\begin{enumerate}
\item{two different initial groups of particles with low and high fractal dimensions 
(such as RLPA for the compact particles and DLPA for the shattered clusters).} 
\item{the flattest kind of particles observed (shattered clusters with an aspect ratio around 0.15) 
could be consistent with compaction of the smallest fractal dimension RLCA and DLCA aggregates 
or with a very large mass loss during collection (>50\%). }
\item{the diversity of morphologies could also originate from a single type of aggregation 
process (such as DLPA) 
but presenting very different cohesive strengths amongst aggregates ($sin(\theta_{0})$ ranging 
from at least 0.1 to 1). This distribution would also present a peak around 0.3 as shown in 
Figure~\ref{fig:effect_Df}, which would be consistent with the peak of the COSIMA distribution
around 0.3 as shown in Figure~\ref{fig:langevin}.}
\item{finally, a fourth process, described in \citet{ellerbroek_footprint_2017}, may be playing 
a role here as well. Experiments show that incoming aggregates may sometimes fragment 
upon impact, leaving some remains sticking to the target in a pyramidal shape
(mass transfer property between 0 and 0.8). }
\end{enumerate}

  The diversity of aspect ratios observed appears consistent with at least 
  two families of aggregates with different $D_f$, which would also be consistent with the 
  GIADA and MIDAS measurements of two dust particles populations with very different fractal 
  dimensions, one being close to 3 and the other around 1.8 
  \citep{fulle_fractal_2017, mannel_fractal_2016}). 
  Variations in both the cohesive strength of the particles and the speed of collection
  may play a role in the continuity of the higher aspect ratio range (>0.3)
  detected by COSIMA.
  Alternatively, this could also mean that the initial low fractal dimensions have been 
  somewhat altered by internal processes, such as compaction, or 
  temperature alteration, such as sintering, which may  have happened 
  during the evolution of the cometary nucleus, especially on its surface. 

\subsection{Comparison with collision experiments}

  In the work of \citet{ellerbroek_footprint_2017}, laboratory simulations of impacts of aggregates
  simulating the particle collection procedure of Rosetta were presented. 
  The aggregates were formed by aggregation of irregular polydisperse SiO$_2$ particles
  with density around \SI{2.6}{\kilo\gram\per\meter\cubed} and a size range 
  of 0.1 to \SI{10}{\micro\meter}.
  The final aggregates have porosities around $65\% \pm 5\%$ and low compressive 
  strength between \SIlist{1e4;1e6}{\pascal}. 
  The aggregates were then accelerated by electrostatic forces towards a collecting 
  plane where the collision was filmed and the resulting flattened footprint imaged and analyzed.  
  The velocity of impact ranges from about \SIrange{1}{6}{\meter\per\second}.
  
  The footprints obtained represent the diversity of morphologies that were acquired
  by the COSIMA instrument. At very low velocities of around \SI{1}{\meter\per\second}, 
  the aggregates either stick directly to the surface, 
  similar to the compact COSIMA particle type, 
  or they may bounce from the surface, 
  leaving a very flat footprint with mostly unconnected fragments, possibly 
  morphologically similar to the shattered cluster COSIMA type of particles. 
  As velocities are increased from \SIrange{2}{6}{\meter\per\second}, 
  the particles mostly stick to the surface
  and fragmentation occurs, leading to footprints morphologically similar 
  to COSIMA rubble piles or glued  clusters.
  
  In this laboratory work, all morphologies were generated using 
  only a change in the impact velocity and impactor size, and similarities
  could be seen between the footprints of the particles that were 
  obtained on the collecting surface and the morphologies measured 
  by COSIMA. 
  The simulations presented in our work allow us to generate 
  similar conditions of flattening by varying the velocity and the 
  particles sizes. However, in our simulations,
  we can also modify the initial impacting particle morphology 
  and study its effect on the flattening of the aggregates. 
  This allows us to explore an extended set of parameters 
  compared with the laboratory experiments, and we have shown that 
  it is also possible to  generate  the measured footprint morphology 
  by considering different initial fractal dimensions of the impacting 
  particles, as discussed above. 
  It would be of interest to study in the laboratory how very 
  porous particles behave when subjected to the type 
  of collection that happened during the Rosetta mission
  to confirm our analysis. 
    
\subsection{Possible analysis of MIDAS data}

  A planned future study aims to investigate whether these results 
  are also valid for MIDAS particles. The aspect ratios of dust particles 
  collected by MIDAS should be calculated and their distribution reviewed. 
  It will be of great interest if the distribution falls in different groups, and if 
  they match those found in the simulation and with COSIMA particles. 
  As MIDAS particles are one order of magnitude smaller than those of 
  COSIMA, this will allow us to understand how the initial structures of dust 
  particles of comet 67P might look and if they remain similar over 
  the \SIrange{1}{100}{\micro\meter} size range.

\section{Conclusions}

  In this work, we have shown that simple numerical simulations of aggregate flattening 
  can be used to infer the initial properties of particles collected by COSIMA on-board Rosetta.
  The diversity of aspect ratios measured in COSIMA images appears consistent with
  several hypotheses on the initial properties of the collected particles. 
  \begin{enumerate}
  \item{It could be explained by at least two families of aggregates with different 
  fractal dimensions $D_f$. A mixture of some compact particles 
  with fractal dimensions close to 2.5-3 together with some  fluffier 
  ones with fractal dimensions <2 would also be consistent with the observations
  and the measurements made by GIADA and MIDAS
  \citep{fulle_fractal_2017,mannel_fractal_2016}.}
  \item{Alternatively, the distribution of morphologies seen by COSIMA could originate from 
  a single type of aggregation process, such as DLPA ($D_f \approx 2.5$)
  but presenting a large range of cohesive strengths or collection velocities. 
  This distribution would be consistent with a maximum at an aspect ratio around 0.3 as observed
  on the COSIMA typology~\citep{langevin_typology_2016}.}
  \end{enumerate}
  Furthermore, variations in cohesive strength and velocity may play a role in 
  the higher aspect ratio range detected by COSIMA (>0.3).
  Our work allows us to explain the particle morphologies observed by 
  COSIMA and those generated by the laboratory experiments of 
  \citet{ellerbroek_footprint_2017} in a consistent framework. 
  Taken together with the observations made by GIADA and MIDAS on 
  Rosetta, our simulations 
  seem to favor an interpretation based on two different families of dust 
  particles with significantly distinct fractal dimensions ejected from 
  the cometary nucleus.

\begin{acknowledgements}
The authors acknowledge two anonymous referees for 
their positive evaluation and constructive comments.
The authors acknowledge support from Centre National d'Etudes Spatiales 
(CNES) in the realization of instruments devoted to space exploration 
of comets and in their scientific analysis. T.M. acknowledges funding by the
Austrian Science Fund FWF P 28100-N36.
\end{acknowledgements}

%
%

\bibliographystyle{aa} 
\bibliography{2018_67P_agg_biblio} 

\end{document}